# On Extracting Thermal Parameters and Scenario in High-Energy Collisions

Ting-Ting Duan[1,a)], Sahanaa Büriechin[1,b)], Hai-Ling Lao[2,c)], Fu-Hu Liu[1,d)], Khusniddin K. Olimov[3,4,e)]

[1]*Institute of Theoretical Physics, State Key Laboratory of Quantum Optics Technologies and Devices & Collaborative Innovation Center of Extreme Optics, Shanxi University, Taiyuan 030006, China*

[2]*Department of Science Teaching, Beijing Vocational College of Agriculture, Beijing 102442, China*

[3]*Laboratory of High Energy Physics, Physical-Technical Institute of Uzbekistan Academy of Sciences, Chingiz Aytmatov Str. 2b, Tashkent 100084, Uzbekistan*

[4]*Department of Natural Sciences, National University of Science and Technology MISIS (NUST MISIS), Almalyk Branch, Almalyk 110105, Uzbekistan*

**Abstract:** The inconsistent thermal parameters derived from various models in high-energy collisions are examined. A comprehensive literature review suggests model-independent parameters to address these inconsistencies, based on the average transverse momentum $\langle p_T \rangle$ and root-mean-square transverse momentum $\sqrt{\langle p_T^2 \rangle}$. The relevant parameters include: initial temperature $T_i \approx \sqrt{\langle p_T^2 \rangle / 2}$, effective temperature $T = \langle p_T \rangle / 2$, kinetic freeze-out temperature $T_0 = \langle p_T \rangle / 6.14$, and average transverse velocity $\beta_T = \langle p_T \rangle / 2\langle m \rangle$, where $\langle m \rangle$ is the average mass of moving particles in the emission source's rest frame. Alternatively, $T_0$ can be seen as the intercept in the linear relationship between $T$ and $m_0$, while $\beta_T$ represents the slope between $\langle p_T \rangle$ and $\langle m \rangle$ (with $m_0$ being the rest mass of a specified particle). Our findings show that these four parameters increase in central collisions, within central rapidity regions, at higher energies, and in larger collision systems. As collision energy rises, excitation functions for all four parameters increase rapidly below approximately 7.7 GeV but slowly above this threshold. At energies greater than 39 GeV, fluctuations appear in these trends with only minor changes observed in their growth rates. This work also reveals a mass-dependent multi-temperature scenario related to both initial states and kinetic freeze-out processes.

a) 202312602001@email.sxu.edu.cn

b) 202201101236@email.sxu.edu.cn

c) hailinglao@163.com ; hailinglao@pku.edu.cn

d) Correspondence: fuhuliu@163.com ; fuhuliu@sxu.edu.cn

e) Correspondence: khkolimov@gmail.com ; kh.olimov@uzsci.net

## 1. Introduction

High-energy collisions are a key research area in modern physics, generating hot and dense systems that undergo hadronization [1]. Currently, these collisions primarily occur at the Brookhaven National Laboratory (BNL) in the USA, with energy levels ranging from several GeV to 200 GeV, and at the European Organization for Nuclear Research (CERN) in Switzerland, which operates between several GeV to nearly 20 GeV and from several TeV to over 10 TeV [2-5]. Additionally, lower-energy related collisions are expected soon at the Helmholtzzentrum für Schwerionenforschung GmbH (GSI) in Germany and the Joint Institute for Nuclear Research (JINR) in Russia [6, 7].

In high-energy collision final states, numerous particles are produced. Some particles appear throughout the entire process—from initial to final state—while the final state also signifies kinetic freeze-out during system evolution. When extracting thermal parameters like kinetic freeze-out temperature ($T_0$) and average transverse flow velocity ($\beta_T$), different particle types contribute variably based on their transverse momentum ($p_T$).

It is widely accepted that most particles are thermally produced via soft processes in the low-$p_T$ region [8-12]. In contrast, a small number of particles arise directly from hard processes in the initial state across all $p_T$ regions. Intermediate-state particles fall between these two categories, distributed across low- and intermediate-$p_T$ regions. Thus, low-$p_T$ particles significantly influence parameters while high-$p_T$ ones have minimal impact; intermediate-$p_T$ contributions are moderate.

Based on varying fit ranges for the same $p_T$ spectrum, parameters extracted from identical models or distributions may differ, not to mention the discrepancies from using different spectra and models. Specifically, applying a blast-wave model across diverse low-$p_T$ fit ranges [13-16], along with other models for high-$p_T$ regions, has led to observations of particle production that can show single, double, or multiple $T_0$ values [12, 13, 17]. This indicates possible scenarios involving one or more kinetic freeze-out processes in high-energy collisions [18-20].

When extracting temperature parameters using the blast-wave model [13-16], confining each $p_T$ fit range to narrow regions corresponding to specific particle types can yield a singular scenario. In contrast, distinct temperature parameters for strange and non-strange particles could suggest a dual scenario. Maximizing the width of each $p_T$ fit range might lead to multiple scenarios generally dependent on particle mass. Additionally, employing alternative models like the improved Tsallis distribution [19, 21] and intercept-slope methods [15, 22-25] may also result in inconsistent temperatures. Thus, kinetic freeze-out scenarios are inherently influenced by both fit range and model selection.

Furthermore, the question of whether $T_0$ is greater in central versus peripheral collisions; within central compared to forward/backward rapidity regions; at low versus high energies; as well as between small versus large collision systems depends on various modeling factors affecting these results [26, 27]. The existence of positive or negative correlations between $T_0$ and $\beta_T$ is also closely tied to specific modeling approaches.

A method to extract the initial temperature ($T_i$), effective temperature ($T$), $T_0$, and $\beta_T$ using average transverse momentum $\langle p_T \rangle$ and root-mean-square transverse momentum $\sqrt{\langle p_T^2 \rangle}$ across the entire $p_T$ range is proposed in this work. This approach is supported by literature research and yields consistent results across different cases, reinforcing multiple temperature scenarios for particle formation and kinetic freeze-out [18, 19, 28, 29].

The remainder of this minireview article is structured as follows: Section 2 introduces single- and multi-component distributions. Section 3 describes model-independent thermal parameters. A discussion on parameter trends is presented in Section 4. Finally, Section 5 provides a summary and conclusion.

## 2. Single- and multi-component distributions

Experimental distributions related to $p_T$ typically take three forms: invariant yield, unit-density of rapidity ($y$) and $p_T$, as well as density of $p_T$. Let $E$ denote energy, $p$ represent momentum, $m_T$ signify transverse mass for a given particle, and $N$ indicate the number of particles; these three forms can be expressed as $Ed^3N/d^3p = (1/2\pi p_T)(d^2N/dydp_T) = (1/2\pi m_T)(d^2N/dydm_T)$, $d^2N/dydp_T$, and $dN/dp_T$, respectively. If experimental data presents $p_T$-related distributions in alternative formats, one may convert these non-canonical representations into one of the aforementioned three forms.

In certain instances where experimental data exhibits narrow ranges in $p_T$ values, a single-component function may suffice to describe or fit these distributions effectively. For example, in proton-proton, proton-nucleus, and nucleus-nucleus collisions conducted at center-of-mass energies per nucleon pair ($\sqrt{s_{NN}}$) ranging from several GeV to over 10 TeV, one can utilize single-component standard (Bose-Einstein or Fermi-Dirac) distribution to approximate experimental data within the region where $p_T < 2-3\ \mathrm{GeV}/c$, in which the particles are produced in the soft excitation process.

The single-component standard distribution arises from the relativistic ideal gas model. The invariant yield is expressed as [30]

$$E\frac{d^3N}{d^3p} = \frac{gV}{(2\pi)^3}E\left[\exp\left(\frac{E-\mu}{T}\right)+S\right]^{-1}, \tag{1}$$

where $g$ and $\mu$ represent the degeneracy factor and chemical potential of a given particle, respectively. Additionally, $V$ and $T$ denote the volume and effective temperature of the interaction system. Here, $S=-1$ corresponds to the Bose-Einstein distribution, while $S=1$ pertains to the Fermi-Dirac distribution. The unit-density for $y$ and $p_T$ is formulated as

$$\frac{d^2N}{dydp_T} = \frac{gV}{(2\pi)^2}p_T\sqrt{p_T^2+m_0^2}\cosh y\left[\exp\left(\frac{\sqrt{p_T^2+m_0^2}\cosh y-\mu}{T}\right)+S\right]^{-1}, \tag{2}$$

with $m_0$ indicating the rest mass of the specified particle. In Eq. (1), $E$ appears in Eq. (2) in terms of $y$ and $p_T$ via $E = m_T \cosh y$, where $m_T = \sqrt{p_T^2 + m_0^2}$ is defined as the transverse mass. The density for $p_T$ is given by

$$\frac{dN}{dp_T} = \frac{gV}{(2\pi)^2} p_T \sqrt{p_T^2 + m_0^2} \int_{y_{\min}}^{y_{\max}} \cosh y \left[ \exp\left( \frac{\sqrt{p_T^2 + m_0^2}\cosh y - \mu}{T} \right) + S \right]^{-1} dy\,, \quad (3)$$

wherein experimental rapidity bins $[y_{\min}, y_{\max}]$ serve as lower and upper limits for integration over $y$.

In Eqs. (1)-(3), $T$ serves as the sole free parameter while $V$ acts as a normalization constant. Although $\mu$ does not significantly influence fits related to distributions involving $p_T$, it can be derived through either particle ratios or traditional methods. The former approach relies on calculating ratio ($K_j$) between negatively charged particles to positively charged ones alongside source's corresponding temperature ($T_j$) at that moment: $\mu = -(1/2)T_j \ln K_j$ [12, 31-33]. The latter method utilizes three net quantum numbers (net-baryon number $B$, strangeness $S$, and electric charge $Q$) alongside their respective chemical potentials ($\mu_B$, $\mu_S$, and $\mu_Q$): $\mu = B\mu_B + S\mu_S + Q\mu_Q$ [12, 34-40].

It is recognized that a single-component standard distribution may not adequately describe extensive distributions related to $p_T$ observed in experiments; thus, consideration should be given to two- or multi-component standard distributions instead. For an $n_0$-component standard distribution scenario, one can express invariant yield as

$$E\frac{d^3N}{d^3p} = \sum_{i=1}^{n_0} E\frac{d^3N_i}{d^3p} = \sum_{i=1}^{n_0} \frac{gV_i}{(2\pi)^3} E \left[ \exp\left( \frac{E - \mu_i}{T_i} \right) + S \right]^{-1}, \quad (4)$$

where parameters such as $N_i$, $V_i$, $\mu_i$, and $T_i$ correspond to particle count, system volume, particle chemical potential, and source's effective temperature, respectively, associated with the $i$-th component. The unit-density for $y$ and $p_T$ takes form

$$\frac{d^2N}{dydp_T} = \sum_{i=1}^{n_0} \frac{d^2N_i}{dydp_T} = \sum_{i=1}^{n_0} \frac{gV_i}{(2\pi)^2} p_T \sqrt{p_T^2 + m_0^2} \cosh y \times \left[ \exp\left( \frac{\sqrt{p_T^2 + m_0^2}\cosh y - \mu_i}{T_i} \right) + S \right]^{-1}. \quad (5)$$

The density of $p_T$ is given by

$$\frac{dN}{dp_T} = \sum_{i=1}^{n_0} \frac{dN_i}{dp_T} = \sum_{i=1}^{n_0} \frac{gV_i}{(2\pi)^2} p_T \sqrt{p_T^2 + m_0^2} \int_{y_{\min}}^{y_{\max}} \cosh y \times \left[ \exp\left( \frac{\sqrt{p_T^2 + m_0^2}\cosh y - \mu_i}{T_i} \right) + S \right]^{-1} dy. \quad (6)$$

In Eqs. (4)-(6), the free parameters are denoted as $T_i$. Generally, a value of $n_0 = 2$ or 3 is sufficient for fitting the data; thus, an excessively large $n_0$ is unnecessary.

In the preceding discussion, although $n_0$ components may exhibit similar forms of standard distribution with varying parameters due to the inherent similarities, commonalities, and universal characteristics observed in high-energy collisions [41-48], increasing the number of free parameters and normalization constants associated with larger values of $n_0$ ultimately reduces the degrees of freedom (ndof), which deviates from our expectations. Instead, one might consider employing the Tsallis distribution [30, 49-54] to encapsulate multi-component standard distributions.

Within the framework of the Tsallis distribution [30, 49-54], invariant yield, unit-density of $y$ and $p_T$, as well as density of $p_T$ are represented by

$$E\frac{d^3N}{d^3p}=\frac{gV}{(2\pi)^3}E\left\{\left[1+(q-1)\frac{E-\mu}{T}\right]^{\frac{1}{q-1}}+S\right\}^{-q}, \quad (7)$$

$$\frac{d^2N}{dydp_T}=\frac{gV}{(2\pi)^2}p_T\sqrt{p_T^2+m_0^2}\cosh y$$
$$\times\left\{\left[1+(q-1)\frac{\sqrt{p_T^2+m_0^2}\cosh y-\mu}{T}\right]^{\frac{1}{q-1}}+S\right\}^{-q}, \quad (8)$$

and

$$\frac{dN}{dp_T}=\frac{gV}{(2\pi)^2}p_T\sqrt{p_T^2+m_0^2}\int_{y_{\min}}^{y_{\max}}\cosh y$$
$$\times\left\{\left[1+(q-1)\frac{\sqrt{p_T^2+m_0^2}\cosh y-\mu}{T}\right]^{\frac{1}{q-1}}+S\right\}^{-q}dy, \quad (9)$$

respectively. Herein, $q$ denotes the entropy index that characterizes the degree of non-equilibrium within the interaction system. Typically, it holds that $q \geq 1$. The special case where $q = 1$ corresponds to equilibrium conditions under which the Tsallis distribution simplifies into its standard one. When $q$ is much greater than 1, it indicates a high degree of non-equilibrium within the system. In practical analyses, it is often found that $q < 1.25$, suggesting that this value does not represent extreme non-equilibrium but rather indicates approximate equilibrium conditions. In Eqs. (7)-(9), there exist two free parameters, $T$ and $q$.

In certain instances where experimental distributions related to $p_T$ span a wide range, a single or two-component Tsallis distribution may prove insufficient for adequately representing all data points. Consequently, consideration must be given to utilizing a multi-component Tsallis distribution which characterized by fewer components than would be required in a corresponding multi-component standard distribution when fitting identical datasets. The structure of this multi-component Tsallis distribution bears similarity to Eqs. (4)-(6); therefore, we will refrain from reiterating those details here.

Empirically speaking, fitting by the multi-component standard or Tsallis distribution can be achieved using a q-dual distribution [55]. With this context, invariant yield, unit-density of $y$ and $p_T$, and density of $p_T$ are denoted by

$$E\frac{d^3N}{d^3p}=\frac{gV}{(2\pi)^3}E\sum_{k=0}^{\infty}(-S)^k\left[1+(k+1)(q-1)\frac{E-\mu}{T}\right]^{-\frac{q}{q-1}}, \tag{10}$$

$$\frac{d^2N}{dydp_T}=\frac{gV}{(2\pi)^2}p_T\sqrt{p_T^2+m_0^2}\cosh y\sum_{k=0}^{\infty}(-S)^k$$
$$\times\left[1+(k+1)(q-1)\frac{\sqrt{p_T^2+m_0^2}\cosh y-\mu}{T}\right]^{-\frac{q}{q-1}}, \tag{11}$$

and

$$\frac{dN}{dp_T}=\frac{gV}{(2\pi)^2}p_T\sqrt{p_T^2+m_0^2}\int_{y_{\min}}^{y_{\max}}\cosh y\sum_{k=0}^{\infty}(-S)^k$$
$$\times\left[1+(k+1)(q-1)\frac{\sqrt{p_T^2+m_0^2}\cosh y-\mu}{T}\right]^{-\frac{q}{q-1}}dy, \tag{12}$$

respectively. In real analysis, a very large value of $k$ is not necessary. In fact, setting the maximum $k$ to 10 is sufficient for obtaining accurate results.

Naturally, one might consider that the multi-component q-dual distribution is well-suited for fitting the $p_T$-related distributions across an extensive range. The multi-component q-dual distribution shares similar summed forms with Eqs. (4)-(6), which will not be reiterated here. Nevertheless, transitioning from standard distribution to q-dual distribution allows for appropriate single-, two-, or multi-component models to fit the $p_T$-related distributions according to varying width ranges.

The multiple components represent various sources characterized by different excitation degrees or interaction mechanisms. Due to temperature variations among these sources, it becomes possible to measure temperature fluctuations ranging from the lowest temperature source ($T_1$), represented by the first component, up to the highest temperature source ($T_{n_0}$), described by the last (the $n_0$-th) component. The average temperature of these $n_0$ sources can be expressed as

$$T=\sum_{i=1}^{n_0}k_iT_i\,. \tag{13}$$

Here, $k_i=N_i/N=V_i/V$ denotes the contribution ratio or fraction of the $i$-th component relative to all $n_0$ components and adheres to normalization such that $\sum_{i=1}^{n_0}k_i=1$.

To illustrate the fitting effectiveness of the three types of distributions discussed above, as an example, Figures 1(a) and 1(b) present invariant yields of $\pi^+$ and $\pi^-$ produced in $|y|<0.1$ in gold-gold (Au+Au) collisions at $\sqrt{s_{NN}}=200$ GeV. Different symbols represent experimental data measured by the STAR Collaboration [2] across various centrality intervals and scaled by distinct quantities indicated in each panel. The solid, dashed, and dotted curves correspond to results fitted using standard, Tsallis, and q-dual distributions with $S=-1$, respectively [12]. It is evident that all three distributions effectively describe the experimental data.

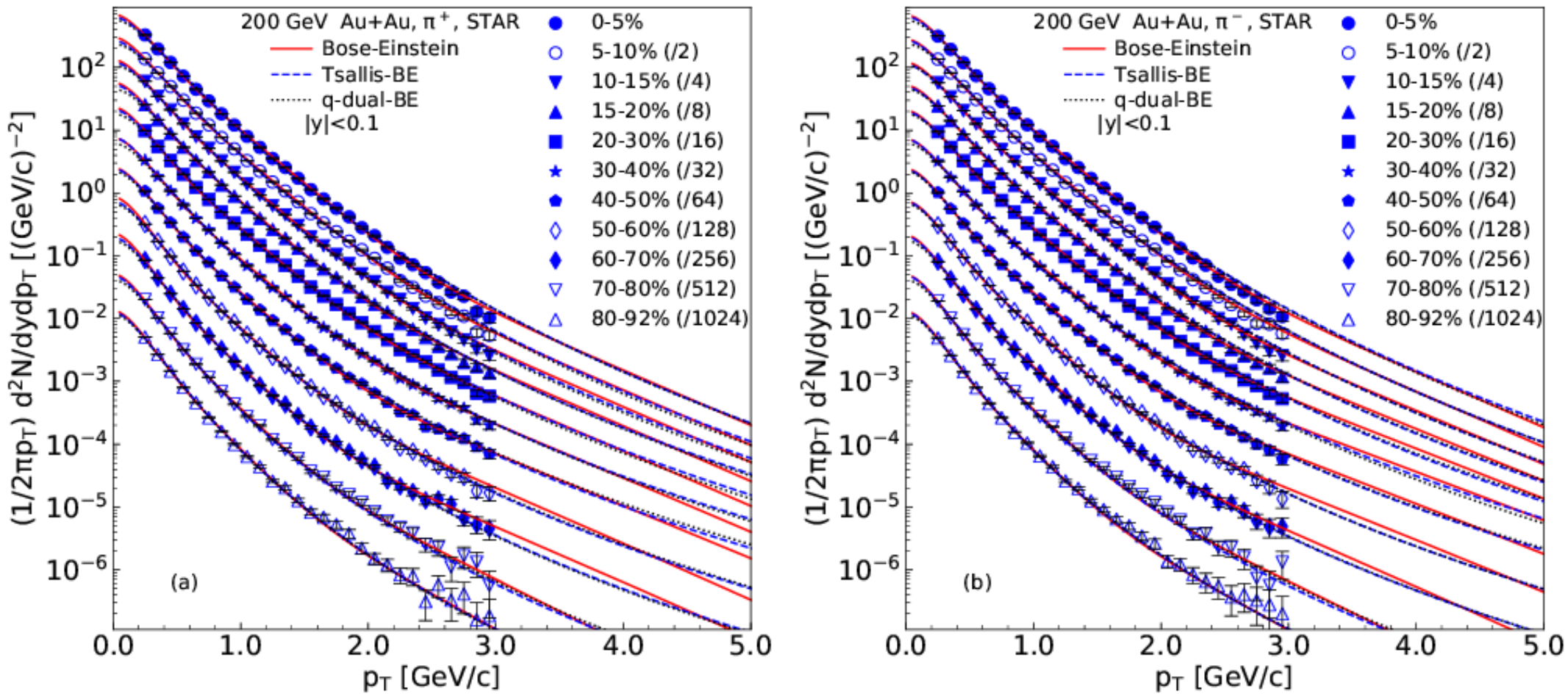

Figure 1. The invariant yields of (a) $\pi^+$ and (b) $\pi^-$ produced in Au+Au collisions at $\sqrt{s_{NN}}=200$ GeV. The symbols represent experimental data measured by the STAR Collaboration [2]. The solid, dashed, and dotted curves correspond to results fitted using standard, Tsallis, and q-dual distributions with $S=-1$, respectively [12].

The effective temperatures $T$ presented in Figure 1 are illustrated in Figure 2 to examine the dependence of $T$ on centrality percentage $C$. Panels (a) and (b) correspond to the results from the spectra of $\pi^+$ and $\pi^-$, respectively. Different symbols represent distinct values of $T$ obtained from various distributions indicated in the panels. It is evident that all three values of $T$ exhibit larger magnitudes in central collisions compared to peripheral collisions. However, these three temperatures show inconsistencies both in absolute values and trends, despite their ability to adequately describe experimental data. The discrepancies arise due to the influence of another parameter, namely the entropy index $q$, within the Tsallis and q-dual distributions.

It is important to note that the three distributions discussed above serve merely as examples for fitting experimental data. Other distributions, such as those derived from the blast-wave model [13-16] and improved Tsallis statistics [19, 21], can also be employed for fitting purposes; however, it is possible that extracted temperatures may yield inconsistent results. Additionally, one might consider the $p_T$-exponential: $dN/p_T dp_T \propto \exp(-p_T/T)$, the $p_T^2$-exponential or $p_T$-Gaussian: $dN/p_T dp_T \propto \exp(-p_T^2/T^2)$, the $p_T^3$-exponential: $dN/p_T dp_T \propto \exp(-p_T^3/T^3)$, the $m_T$-exponential: $dN/m_T dm_T \propto \exp(-m_T/T)$, the Boltzmann: $dN/m_T dm_T \propto m_T \exp(-m_T/T)$, etc. [15], or their superpositions. Here, the effective temperatures $T$ used in these expressions are different.

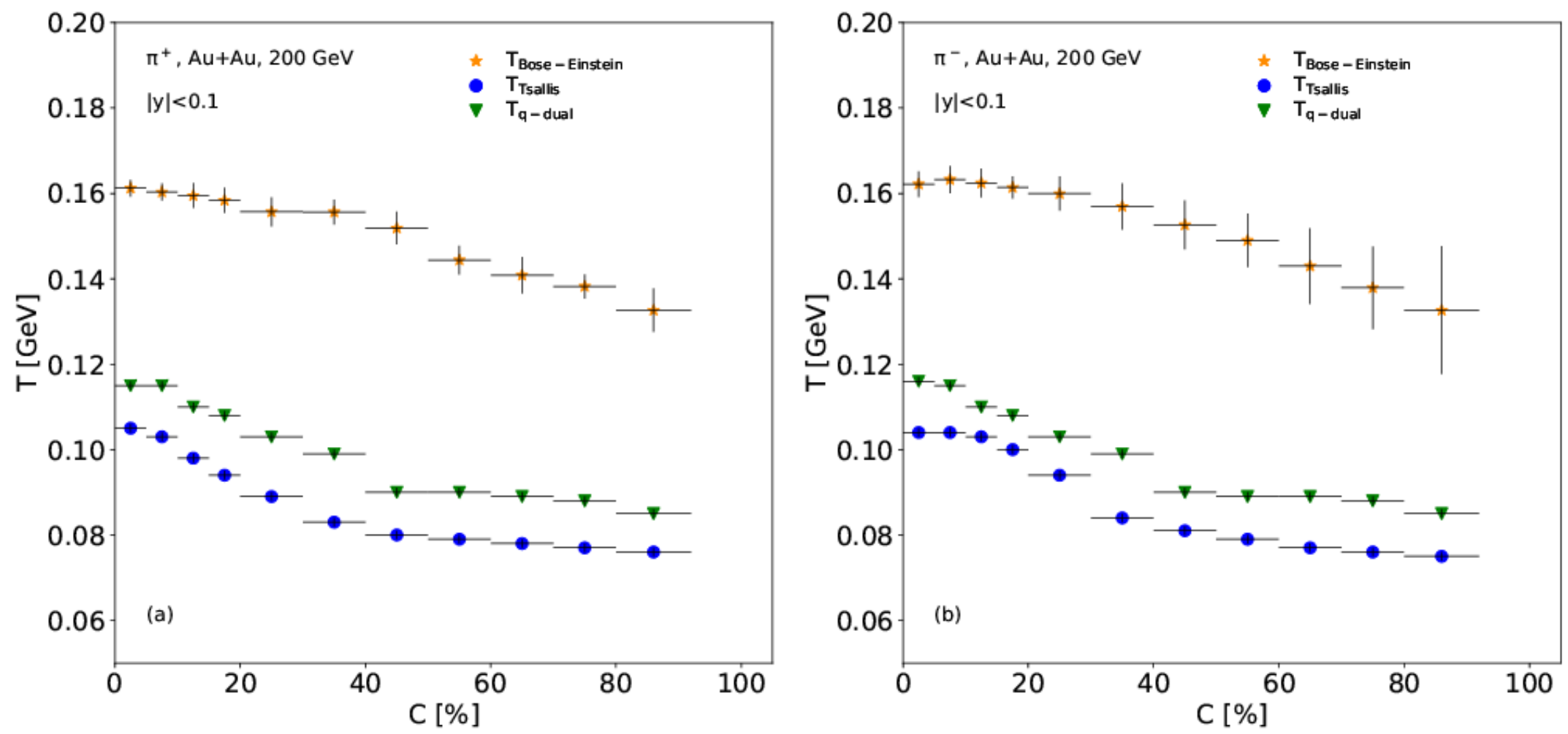

Figure 2. Dependence of $T$ on $C$, where $T$ is derived from standard, Tsallis, and q-dual distributions with $S=-1$, respectively. Panels (a) and (b) present results from $\pi^+$ and $\pi^-$ spectra respectively.

We would like to emphasize that while a variety of distributions can be utilized for fitting spectra related to $p_T$, the most fundamental remains the standard distribution derived from relativistic ideal gas model [30]. Nevertheless, in complex systems such as high-energy collisions, it is unrealistic to expect a single-component standard distribution will suffice. Instead, one should contemplate a multi-component standard distribution capable of accommodating multiple sources with varying degrees of excitation or interaction mechanisms. In real fitting, a standard distribution of two or three components is sufficient.

### 3. Model-independent thermal parameters

In our discussions above, it has been noted that $T$ is dependent on model or distribution choice; this dependency can lead to different absolute values being reported. In particular, since $T$ describes the degree of excitation within an interacting system at kinetic freeze-out, it exhibits greater sensitivity towards model or distribution selection. Furthermore, many existing models or distributions have yet to disentangle contributions arising from thermal motion versus flow effect. Typically, the intensity of thermal motion is characterized by $T_0$, while the magnitude of flow effect is represented by $\beta_T$.

The model-dependent parameters $T_0$ and $\beta_T$ can lead to significantly inconsistent results. For instance, which value of $T_0$ is greater in central versus peripheral collisions? How does it compare between central and forward/backward rapidity regions? What about low-energy (e.g., 10 GeV) versus high-energy (e.g., 200 GeV) collisions? Additionally, how do small collision systems (e.g., proton-proton or proton-nucleus interactions) stack up against large ones (e.g., nucleus-nucleus collisions)? Are $T_0$ and $\beta_T$ directly or inversely proportional? The answers to these questions often yield contradictions based on different models, such as the blast-wave model [13-16] and thermal-related models [56, 57].

Generally speaking, the blast-wave model suggests a smaller value for $T_0$ in central collisions, at high energies, or within larger systems. In contrast, thermal-related models tend to present opposing conclusions. This discrepancy primarily arises from differing assumptions regarding flow profiles and varying fit ranges for $p_T$ distributions [58]. The blast-wave model posits a flow that gradually increases from the inner core to the outer surface with particle-dependent fit ranges for $p_T$. Conversely, thermal-related models assume an invariant flow with broader fit ranges applicable across different particles.

To reconcile these contradictions or inconsistencies, one might consider employing average transverse momentum ($\langle p_T \rangle$) or root-mean-square transverse momentum ($\sqrt{\langle p_T^2 \rangle}$) as representations of thermal parameters. For any given set of data related to $p_T$, both $\langle p_T \rangle$ and $\sqrt{\langle p_T^2 \rangle}$ are independent of specific models or distributions; they can even be derived directly from empirical data. Before delving into the relationship between thermal parameters and either measure $\langle p_T \rangle$ or $\sqrt{\langle p_T^2 \rangle}$, we will first outline their respective representations based on the three distributions related to $p_T$ mentioned above, as shown below.

Whether it is a single component or multiple components, from the distribution that can fit the relevant data of $p_T$, one can conveniently obtain $\langle p_T \rangle$ as

$$\langle p_T \rangle = \int_0^{p_{T\max}} \int_{y_{\min}}^{y_{\max}} p_T^2 E \frac{d^3N}{d^3p} dy dp_T \Big/ \int_0^{p_{T\max}} \int_{y_{\min}}^{y_{\max}} p_T E \frac{d^3N}{d^3p} dy dp_T \,, \tag{14}$$

$$\langle p_T \rangle = \int_0^{p_{T\max}} \int_{y_{\min}}^{y_{\max}} p_T \frac{d^2N}{dy dp_T} dy dp_T \Big/ \int_0^{p_{T\max}} \int_{y_{\min}}^{y_{\max}} \frac{d^2N}{dy dp_T} dy dp_T \,, \tag{15}$$

and

$$\langle p_T \rangle = \int_0^{p_{T\max}} p_T \frac{dN}{dp_T} dp_T \Big/ \int_0^{p_{T\max}} \frac{dN}{dp_T} dp_T \tag{16}$$

if the invariant yield, unit-density of $y$ and $p_T$, and density of $p_T$ are used, respectively. In particular, if the probability density function, $f(p_T) = (1/N)(dN/dp_T)$, is used, one has

$$\langle p_T \rangle = \int_0^{p_{T\max}} p_T f(p_T) dp_T \Big/ \int_0^{p_{T\max}} f(p_T) dp_T = \int_0^{p_{T\max}} p_T f(p_T) dp_T \tag{17}$$

due to the denominator in Eq. (17) being 1.

Go a step further, one has $\sqrt{\langle p_T^2 \rangle}$ in different forms,

$$\sqrt{\langle p_T^2 \rangle} = \sqrt{\int_0^{p_{T\max}} \int_{y_{\min}}^{y_{\max}} p_T^3 E \frac{d^3N}{d^3p} dy dp_T \Big/ \int_0^{p_{T\max}} \int_{y_{\min}}^{y_{\max}} p_T E \frac{d^3N}{d^3p} dy dp_T} \,, \tag{18}$$

$$\sqrt{\langle p_T^2 \rangle} = \sqrt{\int_0^{p_{T\max}} \int_{y_{\min}}^{y_{\max}} p_T^2 \frac{d^2N}{dy dp_T} dy dp_T \Big/ \int_0^{p_{T\max}} \int_{y_{\min}}^{y_{\max}} \frac{d^2N}{dy dp_T} dy dp_T} \,, \tag{19}$$

$$\sqrt{\langle p_T^2 \rangle} = \sqrt{\int_0^{p_{T\max}} p_T^2 \frac{dN}{dp_T} dp_T \Big/ \int_0^{p_{T\max}} \frac{dN}{dp_T} dp_T} \,, \tag{20}$$

and

$$\sqrt{\langle p_T^2 \rangle} = \sqrt{\int_0^{p_{T\max}} p_T^2 f(p_T) dp_T \Big/ \int_0^{p_{T\max}} f(p_T) dp_T = \int_0^{p_{T\max}} p_T^2 f(p_T) dp_T} \,. \qquad (21)$$

If $T$ can be directly obtained from $\langle p_T \rangle$, it is possible to derive a model-independent $T$. If $T$ is proportional to $\langle p_T \rangle$ with a proportionality coefficient of 1, then we have $T = \langle p_T \rangle$. This definition is indeed model-independent. Considering that two participant partons from the projectile and target mainly taking part formation of a special particle, we have a revised effective temperature $T = \langle p_T \rangle/2$ at the parton level.

Regarding $T_0$, a straightforward approach is to assume that $T_0 = \langle p_T \rangle/\kappa$, where $\kappa$ is a constant yet to be determined. In thermal-related models at particle level, one finds that $\kappa = 3.07$ [56, 57]. Similar to the consideration for $T$, we have $\kappa = 6.14$ at the parton level, where $1/2$ is introduced. Considering that thermal motion contributes only $\langle p_T \rangle/2$ to $\langle p_T \rangle$, the remaining contribution $\langle p_T \rangle/2$ to $\langle p_T \rangle$ should arise from flow effect. It follows that $\beta_T = \langle p_T \rangle/2\langle m \rangle$, where $\langle m \rangle = m_0 \langle \gamma \rangle$ is the average mass (energy), and $\langle \gamma \rangle$ is the average Lorentz factor of moving particles in the rest frame of the emission source. It is noted that $T_0/\beta_T = m_0\langle \gamma \rangle/3.07$. Previous studies indicate that $\langle \gamma \rangle$ is less than 0.5% of the Lorentz factor for beam ions producing light flavor particles in TeV energy collisions [28]. Since both parameters, $T_0$ and $\beta_T$, discussed here are solely dependent on $\langle p_T \rangle$ derived from data, they are considered model-independent.

To our knowledge, there has been no discussion in existing literature regarding the relationship between $T$ ($T_0$) and $\sqrt{\langle p_T^2 \rangle}$. Instead, research has focused on examining the connection between initial temperature $T_i$ and $\sqrt{\langle p_T^2 \rangle}$ through color string percolation model [59-61]. According to this model, $T_i = \sqrt{\langle p_T^2 \rangle / 2F(\xi)}$, where $F(\xi) \approx 0.6 - 1$ represents the color suppression factor; a value of 0.6 corresponds to multiple strings with low probability while a value of 1 pertains to single string with high probability.

When considering only single string formed by two partons present within projectile and target ions respectively, we find $T_i \approx \sqrt{\langle p_T^2 \rangle / 2}$, which results in an approximate maximum reduction of 30% compared with $T_i = \sqrt{\langle p_T^2 \rangle / 2F(\xi)}$. In particle formation scenarios, it is most likely for two partons to predominantly participate in collisions, which may form single string. As an estimation, it can be accepted that $T_i \approx \sqrt{\langle p_T^2 \rangle / 2}$. Although this result arises from color string percolation model [59-61], it remains solely data-dependent; thus, indicating that the use of $T_i$ here maintains its model independence.

Although we have obtained expressions for $T_i$, $T$, $T_0$, and $\beta_T$, alternatively, $T_0$ can be interpreted as the intercept in the linear relationship between $T$ and $m_0$, while $\beta_T$ can be viewed as the slope in the linear relationship between $\langle p_T \rangle$ and $\langle m \rangle$. For unification purposes in this work, $\langle p_T \rangle$ and $\sqrt{\langle p_T^2 \rangle}$ are expressed in units of $\mathrm{GeV}/c$. Other quantities are measured as follows: $T_i$, $T$, and $T_0$ are in GeV, while $\beta_T$ is expressed in $c$ (with $c = 1$ within natural units).

Building upon this framework allows for the extraction of model-independent parameters. As an illustration, Figure 3 depicts the dependence of newly defined $T$ on $C$ as derived from Figure 1. Panels (a) and (b) present results corresponding to $\pi^+$ and $\pi^-$ spectra respectively. Distinct symbols represent various values of $T$, each obtained from different distributions indicated within the panels. It is evident that all three values of $T$ exhibit larger magnitudes in central collisions compared to peripheral collisions, a result consistent with observations related to $\langle p_T \rangle$. The three values demonstrate not only agreement in absolute terms but also similar trends across conditions within uncertain range; analogous behavior can be observed for $T_i$, $T_0$, and $\beta_T$, details which will not be elaborated here due to their similarity.

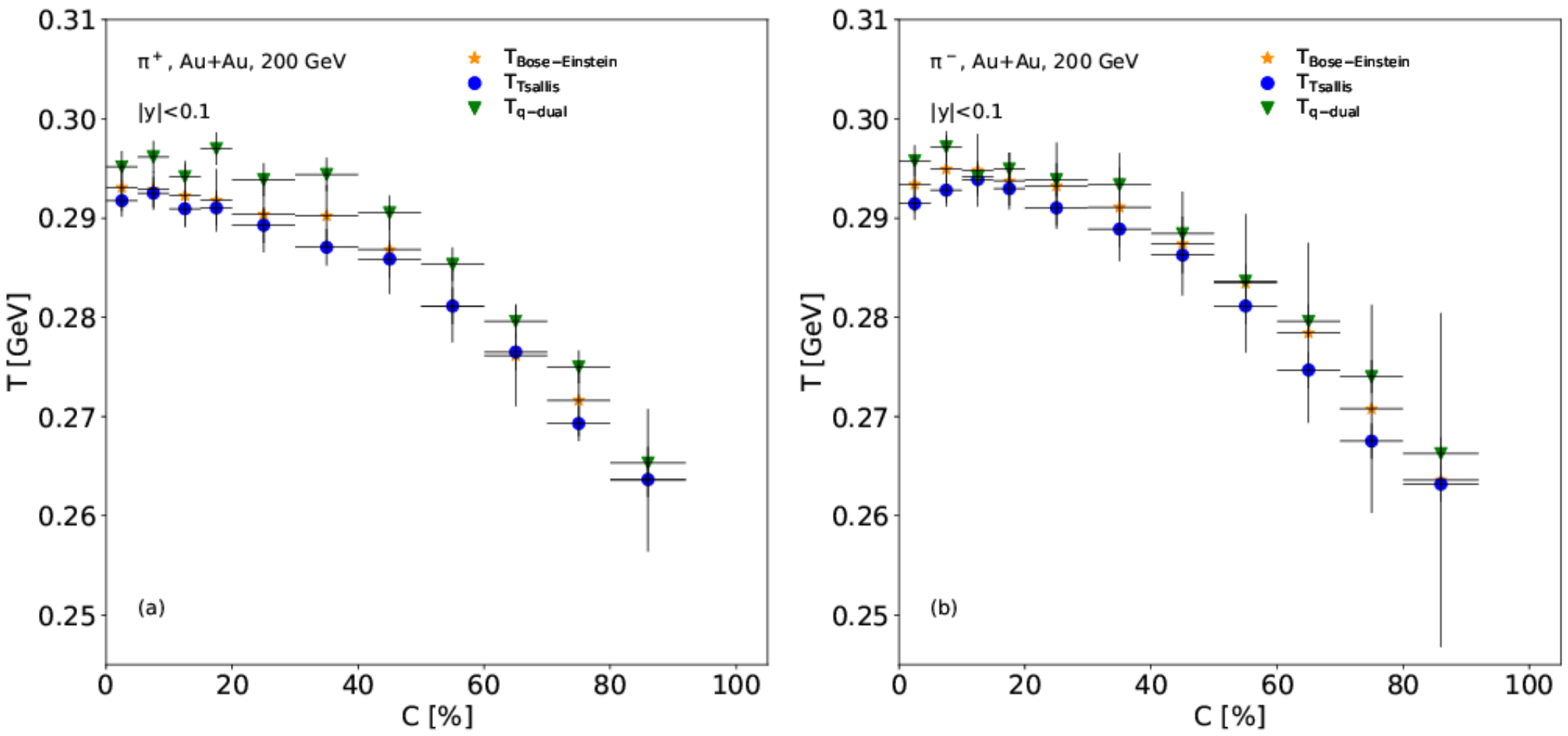

Figure 3. Dependence of $T$ on $C$, where $T = \langle p_T \rangle / 2$; this is based on standard, Tsallis, and q-dual distributions with $S = -1$, respectively. Panels (a) and (b) illustrate results from $\pi^+$ and $\pi^-$ spectra respectively.

### 4. Discussion on parameter trends

We advocate for utilizing these model-independent parameters: $T_i$, $T$, $T_0$, and $\beta_T$. These parameters have been extracted from either $\langle p_T \rangle$ or $\sqrt{\langle p_T^2 \rangle}$. Analyzing trends reported in literature regarding $\langle p_T \rangle$ and $\sqrt{\langle p_T^2 \rangle}$ reveals that both measures tend to be larger in central collisions than peripheral ones; they also increase with collision energy and system size when comparing large systems against small ones under various energy conditions respectively. This observation indicates that all four proposed parameters likewise manifest greater values under circumstances involving central collisions, higher energies, or larger systems, even though some differences may appear minimal [26, 27, 62-65].

As for the rapidity dependence of $\langle p_T \rangle$ and $\sqrt{\langle p_T^2 \rangle}$, if the kinetic energy associated with longitudinal motion is accurately subtracted from the total energy of a given particle, it can yield reasonable results indicating that higher values of $\langle p_T \rangle$ and $\sqrt{\langle p_T^2 \rangle}$ are observed in the central rapidity region compared to the forward/backward rapidity regions. This observation suggests larger values of $T_i$, $T$, $T_0$, and $\beta_T$ in the central rapidity region due to a greater deposition of collision energies.

Utilizing model-independent parameters allows for consistent results across various collisions at different energies. The four parameter values are absolute rather than merely relative because they maintain model independence. Furthermore, since $\langle p_T \rangle$ and $\sqrt{\langle p_T^2 \rangle}$ depend on particle mass, these four parameters exhibit mass dependence. Consequently, this leads to multiple temperature scenarios; specifically, the mass-dependent $T_i$ reflects multiple scenarios of initial state, while the mass-dependent $T_0$ corresponds to multiple scenarios of kinetic freeze-out [18, 19, 28, 29], during both particle formation and kinetic freeze-out processes.

Based on our research experiences [66-70], an examination of the excitation functions for these four parameters reveals a consistent trend. As collision energy increases, these excitation functions rise rapidly within an energy range below approximately 7.7 GeV; beyond this threshold (greater than ~7.7 GeV), their increase becomes more gradual. Subsequently, at several tens of GeV (e.g., around 39 GeV), fluctuations occur in these trends or slight changes in growth rates may be observed.

According to properties associated with Quark-Gluon Plasma (QGP) formed in high-energy collisions, QGP can be classified into three types [71]. A QGP in the usual sense is generated in collisions occurring within an energy range from 7.7 to 39 GeV; a weakly coupled QGP (wQGP) forms at energies below 7.7 GeV; whereas a strongly coupled QGP (sQGP) emerges at energies above 39 GeV. There exists a critical endpoint below 7.7 GeV for the formation of wQGP. It is anticipated that larger volumes of sQGP can be generated at higher energies, such as hundreds of GeV and TeV.

## 5. Summary and conclusions

In summary, this article reviews the inconsistent results regarding thermal parameters derived from various models. To address these discrepancies, we employ model-independent parameters: initial temperature $T_i$, effective temperature $T$, kinetic freeze-out temperature $T_0$, and average transverse flow velocity $\beta_T$. These parameters are determined based on either the average transverse momentum $\langle p_T \rangle$ or the root-mean-square transverse momentum $\sqrt{\langle p_T^2 \rangle}$. Specifically, we propose using: $T_i \approx \sqrt{\langle p_T^2 \rangle/2}$, $T = \langle p_T \rangle/2$, $T_0 = \langle p_T \rangle/6.14$, and $\beta_T = \langle p_T \rangle/2m_0\langle \gamma \rangle$, in data analyses pertaining to high-energy collisions; some of these relationships are drawn from existing literature. This proposal leads to $T_0 = (m_0 \langle \gamma \rangle/3.07)\beta_T$. Flexibly, $T_0$ can be regarded as the intercept in the linear relation between $T$ and $m_0$, while $\beta_T$ can be regarded as the slope in the linear relation between $\langle p_T \rangle$ and $\langle m \rangle$. By implementing this approach, completely consistent results can be achieved.

To conclude, our analyses indicate that the model-independent parameters in central collisions exceed those observed in peripheral collisions. Additionally, measurements within the central rapidity region surpass those found in the forward/backward rapidity regions.

Furthermore, values at high energy (e.g., 200 GeV) are greater than those at low energy (e.g., 10 GeV), and measurements from large collision systems (i.e., nucleus-nucleus collisions) are higher than those from small collision systems (i.e., proton-proton and proton-nucleus collisions). As collision energy increases, excitation functions for all four parameters rise rapidly within a range smaller than approximately 7.7 GeV but increase more slowly beyond this threshold. At higher energies (>39 GeV), fluctuations occur in the trends of these four excitation functions or exhibit slight changes in their growth rates.

In this proposal, while $\langle p_T \rangle$ and $\sqrt{\langle p_T^2 \rangle}$ are independent of any specific model, they demonstrate dependence on particle mass. These dependencies result in $T_i$ exhibiting a mass-dependent multi-scenario for the initial state, whereas $T_0$ and $\beta_T$ reveal a mass-dependent multi-scenario for kinetic freeze-out. The formation and emission of various types of particles do not occur simultaneously; generally, the formation and emission of massive particles precede those of lighter particles during the evolution of the collision system. However, instances where these processes interweave, either occurring earlier or later, can also arise. It is challenging to definitively ascertain whether a single- or two-temperature scenario characterizes both the initial state and kinetic freeze-out; nonetheless, it can be concluded that there exists a mass-dependent multi-temperature scenario.

**Data Availability**

Data sharing is not applicable to this article as no data sets were generated during the current study. The data used to support the findings of this study and some outcome or conclusive statements are included within the article and are cited at relevant places within the text as references.

**Ethical Approval**

The authors declare that they are in compliance with ethical standards regarding the content of this paper.

**Disclosure**

The funding agencies have no role in the design of the study; in the collection, analysis, or interpretation of the data; in the writing of the manuscript; or in the decision to publish the results.

**Conflicts of Interest**

The authors declare that there are no conflicts of interest regarding the publication of this paper.

**Acknowledgments**

The work of Shanxi Group was supported by the National Natural Science Foundation of China under Grant No. 12147215, the Shanxi Provincial Basic Research Program (Natural Science Foundation) under Grant No. 202103021224036, and the Fund for Shanxi "1331 Project" Key Subjects Construction. The work of K.K.O. was supported by the Agency of Innovative Development under the Ministry of Higher Education, Science and Innovations of the Republic of Uzbekistan within the fundamental project No. F3-20200929146 on analysis of open data on heavy-ion collisions at RHIC and LHC.